\documentclass[preprint]{revtex4}
 \pdfoutput=1 

\usepackage{graphicx}
\usepackage{latexsym}
\usepackage{amsmath, amsthm, amssymb}
 \usepackage{txfonts} 
 \usepackage{comment}
 \usepackage{color}
\usepackage{hyperref}
\usepackage{xcolor}
\definecolor{dark-red}{rgb}{0.4,0.15,0.15}
\definecolor{dark-blue}{rgb}{0.15,0.15,0.4}
\definecolor{medium-blue}{rgb}{0,0,0.5}
\hypersetup{
    colorlinks, linkcolor={dark-blue},
    citecolor={dark-blue}, urlcolor={medium-blue}
}

\usepackage{feynmp-auto}
\usepackage[latin1]{inputenc}
\usepackage[english]{babel}
\usepackage[T1]{fontenc}
\usepackage{bbm}
\usepackage{graphicx}
\usepackage{esint}
\usepackage{pdflscape}
\usepackage{slashed}
\usepackage{float}
\usepackage{color}
\usepackage{csquotes}

\usepackage{latexsym}
\usepackage{amsmath, amsthm, amssymb}
 \usepackage{txfonts} 
 \usepackage{comment}
 \usepackage{color}
\usepackage{hyperref}
\usepackage{xcolor}
\definecolor{dark-red}{rgb}{0.4,0.15,0.15}
\definecolor{dark-blue}{rgb}{0.15,0.15,0.4}
\definecolor{medium-blue}{rgb}{0,0,0.5}
\hypersetup{
    colorlinks, linkcolor={dark-blue},
    citecolor={dark-blue}, urlcolor={medium-blue}
}

\newcommand{\dif}[1]{\ensuremath{\medspace \mbox{d} #1}}
\newcommand{\pdif}[2]{\ensuremath{\frac{\partial #1}{\partial #2}}}

\newcommand{\m}[1]{\ensuremath{\mathrm{#1}}}

\begin{document}
\title{Dual QED$_3$ at ``$N_F = 1/2$'' is an interacting CFT in the infrared} 

\author{Dietrich Roscher}
\email{droscher@thp.uni-koeln.de}

\author{Emilio Torres}
\email{torres@thp.uni-koeln.de}

\author{Philipp Strack}
\email{strack@thp.uni-koeln.de}

\homepage{http://www.thp.uni-koeln.de/~strack/}

\affiliation{Institut f\"ur Theoretische Physik, Universit\"at zu K\"oln, D-50937 Cologne, Germany}

\date{\today}

\begin{abstract}
We study the fate of weakly coupled dual QED$_3$ in the infrared, that is, 
a single two-component Dirac fermion coupled to an emergent U(1) gauge field, but without 
Chern-Simons term. 
This theory 
has recently been proposed as a dual description of 2D surfaces of certain topological insulators. 
Using the renormalization group, we find that the interplay of gauge fluctuations with 
generated interactions in the four-fermi sector stabilizes an interacting 
conformal field theory (CFT) with finite four-fermi coupling in the infrared. 
The emergence of this CFT is due to cancellations 
in the $\beta$-function of the four-fermi coupling special to ``$N_F = 1/2$''.
We also quantify how a possible ``strong'' Dirac fermion 
duality between a free Dirac cone and dual QED$_3$ would constrain
the universal constants of the topological current correlator of the latter.
\end{abstract}

\maketitle
\tableofcontents

\section{Introduction}
There are several proposals claiming that more than one strongly 
interacting topological phase of matter might be describable by a 
unifying dual theory: a single two-component (composite) 
Dirac fermion coupled to an emergent $U(1)$ gauge field \cite{son15,metlitski15,wang15}:
\begin{align}
\mathcal{L}_{cf} = \bar{\psi}\left(i\slashed{\partial}+\bar{e}\slashed{a}\right)\psi
+ ...\;.
\label{eq:lagrangian}
\end{align}
Here $\bar{\psi}$ are not the fundamental electrons 
but two-component composite fermions.  And $\bar{e}$ is not the physical, electric 
charge but the charge with respect to the emergent gauge field $a_\mu$. 
There is no Chern-Simons term for $a_\mu$ and this is due to a restricted set of large 
gauge transformations. That is, only certain types of electric fluxes are allowed on the 
surface of the topological insulator \cite{metlitski15}. The dots stand 
for terms that are allowed by symmetries to be generated under a 
renormalization group (RG) flow, such as four-fermi interactions and radiative 
corrections to the propagators.

In high-energy physics, Dirac fermions are typically collected in a 4D-representation of the 
fermion spinor algebra; the case of interest in the present paper therefore laxly corresponds to 
``$N_F \equiv N^{(4)}_{\rm f} = 1/2$''. It has not received much attention 
so far due to its novelty and only very recent association to topological condensed 
matter systems. Moreover, this case falls outside the region of validity of the
Vafa-Witten theorem, which states that vector-like gauge theories
cannot dynamically generate fermion masses in $\bar{\psi}\psi$ channels, 
thereby breaking time-reversal and space-reflection symmetry \cite{vafa84}. 
But their proof only holds for a number of two-component fermions
$N^{(2)}_{\rm f} \geq 4$ and even.

More broadly, a Dirac fermion analog to the particle-vortex duality 
of bosons has been proposed \cite{metlitski15} and an explicit construction of a possible 
``strong'' duality of Eq.~(\ref{eq:lagrangian}) to a free Dirac cone of electrically charged electrons 
on the level of partition functions has been put forward in Ref.~\onlinecite{mross15}. 

It is fair to say that whether, and in what form, and on what energy scales this Dirac fermion 
duality holds is an open question with fundamental implications across a variety of physical systems. 
In particular, not much is known about the low-energy fate and ground state phase diagram 
of Eq.~(\ref{eq:lagrangian}), which we will refer to as dual QED$_3$ in what follows.

In this paper, we want to start filling this knowledge gap and we ask if/under what conditions the 
low-energy dynamics of weakly coupled dual QED$_3$, Eq.~(\ref{eq:lagrangian}), remains 
conformal and by which mechanism interactions can generate a mass for the (composite) fermions 
at low energies. To achieve this, we will adapt the symmetry-breaking analysis of $N^{(4)}_{\rm f}$ 
four-component fermions coupled to a $U(1)$ gauge field of Ref.~\onlinecite{braun14} to a single 
two-component $N_{\rm f}^{(2)}=1$ fermion and compute the low-energy fixed-point structure for an initially 
weakly coupled Eq.~(\ref{eq:lagrangian}). The question of (chiral) symmetry-breaking of QED$_3$ 
has been tackled intensely and we refer to the Introduction and Bibliography of 
Refs.~\onlinecite{braun14,GTK16,giombi15} for an overview.

\subsection{Key results and outline of paper}

Our main result is that Eq.~(\ref{eq:lagrangian}), if initially weak-to-moderately coupled, 
flows toward an interacting conformal field theory (CFT) in the infrared 
in which generated four-fermi couplings attain finite values. 
In the condensed matter context of dual QED$_3$ as a surface description 
for topological insulators, this implies that interactions 
preserve the gapless responses and the 
intrinsic topological order also at lowest energies. This result is derived and 
presented in Subsec.~\ref{subsec:topo}. 

In light 
of strong (chiral and vector) symmetry-breaking tendencies for QED$_3$
at single-digit flavor number $N_{\rm f}^{(4)}$ \cite{braun14}, this result 
may seem surprising. On the other hand, the $N_{\rm f}^{(2)} = 1$ case does 
not have the full chiral symmetry to begin with, as it may be viewed to operate 
within one chiral sector. The number of symmetries that can be broken 
is now reduced and there remains essentially only one independent four-fermi coupling ($\lambda$)
of the associated Fierz algebra (see Sec.~\ref{sec:topo}). 
With this, we find that gauge fluctuations never destabilize the four-fermi sector 
toward symmetry-breaking sufficiently strongly; instead, the flow is 
{\it always} attracted toward an infrared stable fixed-point for the four-fermi coupling, 
which preserves the scaling/conformal symmetry of the gauge sector.
Surprisingly, this is due to the absence of a $\lambda^2$ term 
in the $\beta$-function for the four-fermi coupling: the flavor 
trace carries a $\sim (N^{(2)}_{\rm f} - 1) \lambda^2$. This is what 
stabilizes the CFT at $N^{(2)}_{\rm f} = 1$. 
Related cancellations of $\beta$-functions in the single-flavor case 
also appear in the Gross-Neveu model; in particular these also hold at 
higher loop orders \cite{wetzel85,gracey90, luperini91}.

We believe our results are not limited to the RG technique used. 
An $\epsilon$-expansion \cite{DiPietro16} around the weakly 
interacting Gaussian fixed-point in $D=4$ should recover the same physics.


In Sec.~\ref{sec:recap}, we recapitulate how related runaway flows 
of different physical origin have been detected in Ref.~\onlinecite{braun14}.  
We present the Fierz-complete action of $N^{(4)}_{\rm f}$ 4-component Dirac fermions 
from which we project out the $\beta$-functions for a single 
two-component fermion, $N^{(2)}_{\rm f} = 1$ in Sec.~\ref{sec:topo}.

In Sec.~\ref{sec:strong}, we explore consequences of a possible 
strong Dirac fermion duality.  We determine {\it exactly}
the universal constant of the topological current correlator 
of Eq.~(\ref{eq:lagrangian}), an interacting theory, by relating 
it to the electromagnetic response of a free Dirac cone.
Finally, in Subsec.~\ref{subsec:mass} we point out the need 
to include the generated four-fermi coupling (and possibly other 
ingredients) in order to establish 
exponent identities for operator dimensions in compliance 
with the duality. 

In Sec.~\ref{sec:conclu}, we conclude the paper.
Details of a direct derivation of the $\beta$-function 
for the four-fermi coupling $\lambda$ are relegated to two Appendices 
\ref{app:betalambda}, \ref{FermiSelfApp}.

\section{Conformal scaling and its breakdown for QED$_3$ with $N^{(4)}_{\rm f}$ 
4-component fermions}
\label{sec:recap}

In Ref.~\onlinecite{braun14}, $\beta$-functions for QED$_3$ were calculated using 
a 4D-reducible representation of the fermionic spinor fields. The main scope of this work was an 
investigation of the symmetry breaking patterns of QED$_3$, 
including chiral channels, by detection of runaway flows for fermionic couplings caused by fixed point annihilation. 
To set the stage for dual QED$_3$, we now recapitulate the key elements of this analysis.

\subsection{Fierz-complete action}

Based on the bare action of a Maxwell term for the photons coupled to a set of $N_{\rm f}^{(4)}$ flavors 
of 4-component Dirac fermions (flavor index implicit)
\begin{equation}
S = \int\dif^3{x}
\left\{\bar{\psi}\left( i\slashed{\partial} + \bar{e}\slashed{a}\right)\psi
+ \frac{1}{4} F_{\mu\nu}F^{\mu\nu}\right\},
\label{eq:model}
\end{equation}
the following Fierz-complete ansatz for the euclidean, scale ($k$-) dependent effective action 
is sufficient to study symmetry-breaking into the complete set of all possible fermionic 
channels
\begin{equation}
\Gamma_k[\bar{\psi},\psi,a] = \int\dif^3{x}\left\{\bar{\psi}\left(iZ_\psi\slashed{\partial} + \bar{e}\slashed{a}\right)\psi + \frac{Z_a}{4}F_{\mu\nu}F_{\mu\nu} + \frac{Z_a}{2\xi}(\partial_\mu a_\mu)^2 + \frac{\tilde{\bar{g}}}{2N^{(4)}_{\m{f}}}(\bar{\psi}\gamma_{45}\psi)^2 + \frac{\bar{g}}{2N^{(4)}_{\m{f}}}(\bar{\psi}\gamma_\mu\psi)^2\right\}\;.
\label{eq:truncation}
\end{equation}
Here the last two terms $\tilde{\bar{g}}$, $\bar{g}$ are two four-fermi couplings from which all 
possible interaction channels, which can lead to condensation of fermion bilinears, 
can be constructed. $\xi$ is a gauge fixing parameter which will be set to $\xi = 0$ in the following (Landau gauge). 
In total, Eq.~(\ref{eq:truncation}) has 5 running couplings 
($Z_\psi$, $\bar{e}$, $Z_a$, $\tilde{\bar{g}}$ and $\bar{g}$),
which depend on the cutoff scale $k$. We are interested in their evolution 
in the infrared as we take $k\rightarrow 0$.

\subsection{$\beta$-functions}

In the simplest, point-like truncation for the couplings, projected onto the most singular 
point in frequency- and momentum space (the origin at $q=0$), the leading order $\beta$-functions for
the gauge coupling $e^2$ and the two four-fermi couplings $\tilde{g}$, $g$ of Eq.~(\ref{eq:truncation}) are:
\begin{subequations}
\label{4DBetaFuncs}
\begin{align}
\partial_t e^2 &= (\eta_a-1)e^2 \label{GaugeBeta}\\
\partial_t \tilde{g} &= \tilde{g}(1+2\eta_\psi) - l_\psi^1\left(\frac{2N_{\m{f}}^{(4)}-1}{N_{\m{f}}^{(4)}}
\tilde{g}^2 - \frac{3}{N_{\m{f}}^{(4)}}\tilde{g}g - \frac{2}{N_{\m{f}}^{(4)}}g^2\right) - l_{a,\psi}^{1,1}
\left(2\tilde{g}e^2+4ge^2\right) + l_{a,\psi}^{2,1} 2N^{(4)}_{\m{f}}e^4\label{gBeta}\\
\partial_t g &= g(1+2\eta_\psi) + l_\psi^1\left( \frac{1}{N_{\m{f}}^{(4)}}\tilde{g}g + \frac{2N_{\m{f}}^{(4)}+1}{3N_{\m{f}}^{(4)}}g^2\right) - \frac{l_{a,\psi}^{1,1}}{3}\left(4\tilde{g}e^2 - 2ge^2\right) \label{sBeta}\;.
\end{align}
\end{subequations}
Here we have abbreviated the scale-derivative $\partial_t = k\, \partial_k$. 
The set of $\beta$-functions Eq.~(\ref{4DBetaFuncs}) is closed by two anomalous 
dimensions making it 5 equations and 5 couplings to solve.
$\eta_{\psi}$ for the electrons
\begin{equation}
\eta_\psi = -\frac{e^2}{3}\left[2\tilde{m}_{a,\psi}^{1,1}-2m_{a,\psi}^{2,1} \right]
\label{eq:eta_psi}
\end{equation}
turns out to be negative in the regimes of interest and the threshold coefficients here 
take the form
\begin{align}
\tilde{m}_{a,\psi}^{1,1} &= \frac{3}{2} - \frac{1}{6}\eta_\psi - \frac{1}{4}\eta_a
\nonumber\\
m_{a,\psi}^{2,1} &= 1 - \frac{1}{4}\eta_a\;.
\end{align}

The photon anomalous dimension $\eta_a$ is physically caused by decay and recombination into 
electron-positron pairs. It takes the form  
\begin{align}
\label{eq:eta_a}
\eta_a =& N_{\m{f}}^{(4)}e^2\frac{1}{\zeta^2}\int_0^\infty\dif{y}
\Bigg\{
\frac{2}{3}
\frac{\partial_t r_\psi(y)-\eta_\psi r_\psi(y)}
{\sqrt{y}\left[1+r_\psi(y)\right]^3} -
 \frac{1}{2}\int_{-1}^1\dif{x}\frac{\sqrt{y}x^2-\zeta x}{y-2\zeta x\sqrt{y} + \zeta^2}
\\
&\left[
\frac{[\partial_t r_\psi](y)-\eta_\psi
r_\psi(y)}{[1+r_\psi(y)]^2[1+r_\psi(y-2\zeta x\sqrt{y}+\zeta^2)]}
+
\frac{[\partial_t r_\psi](y-2\zeta x\sqrt{y}+ \zeta^2)-\eta_\psi r_\psi(y-2\zeta x\sqrt{y}+\zeta^2)}
{[1+r_\psi(y)][1+r_\psi(y-2\zeta x\sqrt{y}+\zeta^2)]^2}\right]\Bigg\},
\nonumber
\end{align}
which has a finite $\zeta \rightarrow 0$ limit.
We also absorbed explicit $k$-dependences into renormalized gauge and induced 
four-fermion couplings
\begin{equation}
\label{RenPres}
\tilde{g} \equiv \frac{\tilde{\bar{g}}k}{Z_\psi^2}, \quad g \equiv \frac{\bar{g}k}{Z_\psi^2},
\quad e^2 \equiv \frac{\bar{e}^2}{Z_\psi^2 Z_a k}.
\end{equation}
The threshold functions appearing in Eq.~(\ref{4DBetaFuncs}) are (for the linear Litim regulator)
\begin{align}
l_\psi^1 &= \frac{2}{3} - \frac{1}{6}\eta_\psi\\
l_{a,\psi}^{1,1} &= \frac{4}{3} - \frac{1}{6}\eta_\psi - \frac{2}{15}\eta_a\\
l_{a,\psi}^{2,1} &= 2 - \frac{1}{6}\eta_\psi - \frac{4}{15}\eta_a
\end{align}
and are positive in the regimes of interest, that is, the ``RG-corrections''
by the anomalous dimensions are subdominant when compared to the leading term.

The 5 $\beta$-functions Eq.~(\ref{4DBetaFuncs})  and 
Eqs.~(\ref{eq:eta_psi},\ref{eq:eta_a}) have scale-invariant, real-valued solutions
for large enough $N^{(4)}_{\rm f} > N^{(4)}_{\rm f,c}$; these signify a conformal phase.
We now first describe the nature of these conformal fixed-points and subsequently 
explain how the scaling breaks down at $N^{(4)}_{\rm f} = N^{(4)}_{\rm f,c}$.

\subsection{Recap of interacting conformal fixed point for $N^{(4)}_{\rm f} > N^{(4)}_{\rm f,c}$}
\begin{figure}
\begin{minipage}{100mm}
\includegraphics[width=100mm]{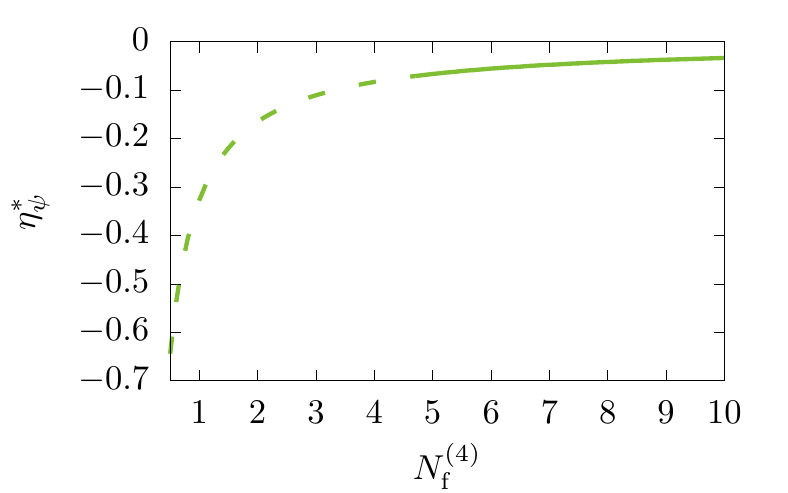}
\end{minipage}
\hspace{5mm}
\begin{minipage}{40mm}
\centering
\begin{tabular}{|c|c|c|}
\hline
$N_{\m{f}}^{(4)}$ & $\eta_\psi^*$ & $e_*^2$\\
\hline
5 & -0.066 & 0.195 \\
6 & -0.055 & 0.163 \\
7 & -0.048 & 0.140 \\
8 &	-0.042 & 0.123 \\
9 &	-0.037 & 0.109 \\
10 & -0.033 & 0.099 \\
\hline
\end{tabular}
\end{minipage}
\caption{Fermion anomalous dimension and value of the gauge coupling
in the conformal phase as a function of the flavor number $N_{\m{f}}^{(4)}$. The dashed line indicates 
the regime below $N_{\m{f,c}}^{(4)}$, where the conformal fixed point becomes unstable and likely 
spontaneous symmetry breaking sets in.}
\label{etaPsiNf}
\end{figure}

Due to charge conservation, the photon anomalous dimension is exactly equal to one
for any value of $N^{(4)}_f>N^{(4)}_{\rm f,c}$:
\begin{align}
\eta^a_\ast = 1\;,
\end{align}
that is, along the line of interacting conformal fixed points corresponding to the conformal phase. 
This follows from Eq.~(\ref{GaugeBeta}). 
Since $\eta_\ast^a$ depends on $e_\ast^2$ itself, this 
fixes the numerical value of the gauge coupling, given in Fig.~\ref{etaPsiNf}, as 
a function of $N^{(4)}_{\rm f}$.
The values of $\eta_\psi^*$ depend on the number of fermion flavors. 
This follows from a solution of the coupled equations for the anomalous dimensions 
Eqs.~(\ref{eq:eta_psi},\ref{eq:eta_a}). 
Its values are given in Fig.~\ref{etaPsiNf} for the linear regulator and $\zeta\rightarrow 0$.
Alternative techniques to access the conformal phase and its exponents and operator 
dimensions are the $1/N_{\rm f}$ expansion, which offers perturbative control for 
sufficiently large $N_{\rm f}$ 
(e.g.: \cite{gracey94,chen93, rantner02, kaul08, pufu14,huh15,chester16, GTK16}), 
and the $\epsilon$-expansion around $d=4$ in the limit $\epsilon\rightarrow 1$ 
(e.g.: \cite{giombi15, DiPietro16, GTK16}).

It is a feature of the $\beta$-functions that the flow of the gauge coupling~\eqref{GaugeBeta} and consequently the 
universal fixed-point values $e_*^2(N_{\rm f})$ shown in Fig.~\ref{etaPsiNf} do not depend on $\tilde{g}$ or $g$. 
At the level of the perturbative Ward identity, this is diagrammatically due to Furry's theorem, i.e., the vanishing of graphs with an 
odd number of external gauge field insertions. 
This decoupling of the gauge flow from the fermion sector permits a simplified analysis of symmetry 
breaking patterns. $e^2$ may be viewed as an external parameter for the fermionic flow 
equations~\eqref{gBeta} and~\eqref{sBeta}. Here we have taken the fermionic couplings to be 
not fundamental in the UV, $\tilde{g}_{k=\Lambda} = g_{k=\Lambda} = 0$. They first 
need to be generated by gauge field fluctuations. 

We now recapitulate how to detect symmetry breaking from the flows of 
Eqs.~(\ref{4DBetaFuncs},\ref{eq:eta_psi},\ref{eq:eta_a}).

\subsection{Breakdown criterion of conformal scaling at $N^{(4)}_{\rm f,c}$}
\label{subsec:breakdown}

For sufficiently large $N_{\rm f}^{(4)}>N^{(4)}_{\rm{f,c}}$, the initial values, 
$g_{k=\Lambda} = \tilde{g}_{k=\Lambda} = 0$ lie in the basin of attraction of a 
conformal fixed point. Note that 
in general there are four fixed point solutions, only one of which is infrared attractive. 
This is the conformal fixed-point at finite $e_{k\rightarrow 0 }^2=e^2_\ast$ (labelled 
as $\mathcal{O}$).

\begin{figure}
\centering
\includegraphics[width=125mm]{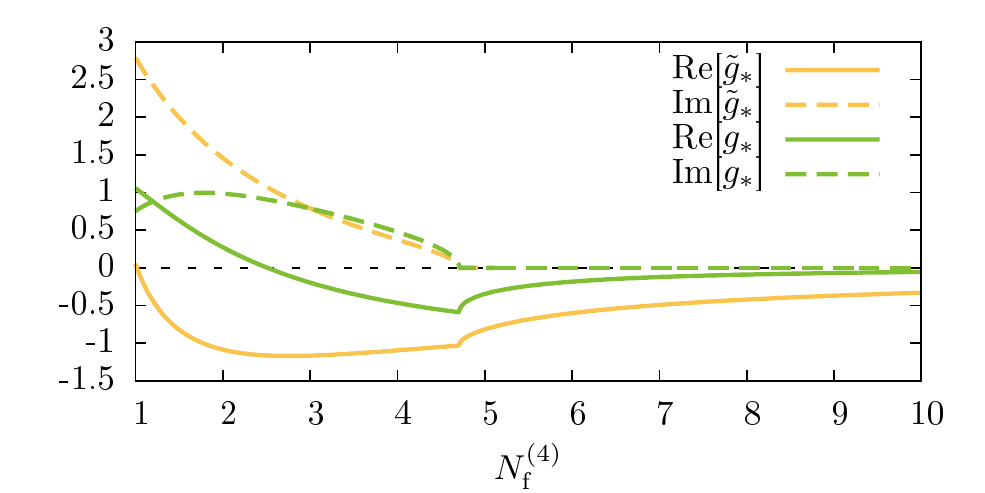}
\caption{Fixed point values of the fermionic couplings $\tilde{g}$ and $g$ at the scale-invariant/conformal fixed point $\mathcal{O}$ for a range of $N^{(4)}_{\m{f}} > N^{(4)}_{\rm f,c} $. The gauge coupling is set to the respective fixed point value $e_*^2(N_{\m{f}}^{(4)})$. Finite imaginary parts indicate spontaneous symmetry-breaking. Here this happens below $N^{(4)}_{\rm f,c} = 4.7$.}
\label{ConfEvo}
\end{figure}

For $N_{\rm f}^{(4)}\leq N^{(4)}_{\rm{f,c}}$, however, the four-fermi couplings at $\mathcal{O}$ start 
developing imaginary parts, which is indicative of spontaneous 
symmetry breaking and the phase boundary between the conformal phase 
and a phase with spontaneously broken symmetry.

In Fig.~\ref{ConfEvo}, we plot the fixed-point values of $g$ and $\tilde{g}$ 
in the complex plane for varying $N_{\rm f}^{(4)}$. We observe that at 
$N_{\rm f}^{(4)}\leq N^{(4)}_{\rm f,c} = 4.7$ the couplings develop imaginary parts.
This estimate is coincidentally close to a recent computation from 
the F-theorem and a resummed $\epsilon$-expansion 
$N^{(4)}_{\rm f,c} \approx 4.4$ \cite{giombi15} and another recent estimate from the $\epsilon$-expansion 
at $N^{(4)}_{\rm f,c} \approx 4.5$ \cite{DiPietro16}.

An explicit solution of the 5 coupled flow equations as a function of 
$k$ confirms this picture: for $N^{(4)}_{\m{f}} \leq 4.7$, the four-fermi couplings 
diverge at some finite scale $k_{\m{sb}}$. These runaway flows 
indicate that fluctuations in one, or several, fermion bilinear channels 
become so strong that one, or a combination, of bilinears 
are likely to condense and spontaneously break the conformal symmetry.

\begin{figure}
\centering
\begin{tabular}{ccc}
(a): $N_{\m{f}}^{(4)} = 6 > N_{\m{f,c}}^{(4)}$ && 
(b): $N_{\m{f}}^{(4)} = 4 < N_{\m{f,c}}^{(4)}$\\
\begin{minipage}{80mm}
\includegraphics[width = 80mm]{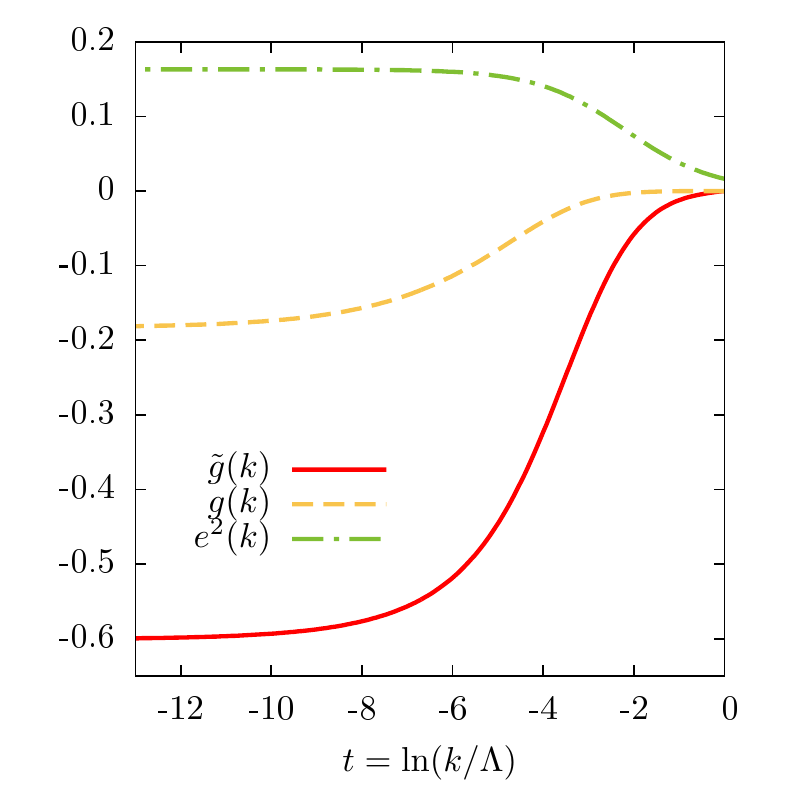}
\end{minipage}
&
&
\begin{minipage}{80mm}
\includegraphics[width = 80mm]{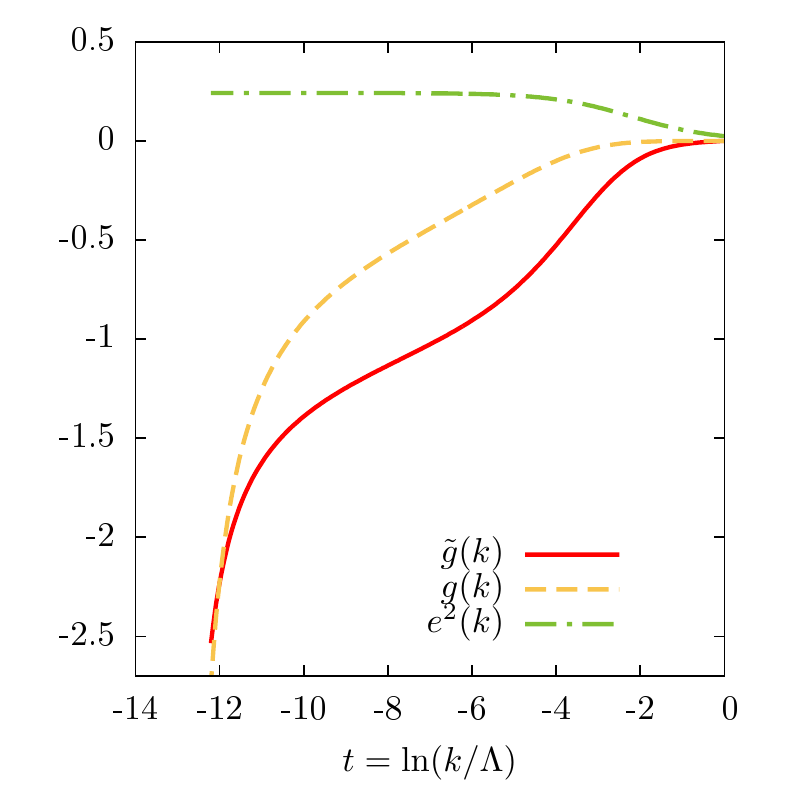}
\end{minipage}
\end{tabular}
\caption{Explicit solutions of the flow equations in the conformal phase (a) and 
in the phase with putatively broken symmetry (b). In the second case, the fermionic flow diverges at $t_{\m{sb}} = -12.9$.}
\end{figure}

\section{Dual QED$_3$ with a single 2-component fermion $N^{(2)}_{\rm f} =1$}
\label{sec:topo}

For 2D surfaces of certain 
3D topological insulators and possibly other topological phases, we are interested in QED$_3$ with a
{\it single} two-component (composite) fermion coupled to 
an emergent $U(1)$ gauge field. Therefore, the 4D representation 
for the fermionic spinors is not applicable and we need to 
move to the 2D-representation and set $N^{(2)}_{\rm f} = 1$. 

\noindent 
To that end, consider the two linearly independent fermionic interactions in the 2D representation defined in the appendix of 
Ref.~\onlinecite{braun14}:
\begin{equation}
\left(\bar{\psi}^a \gamma_{45}\psi^a\right)^2_{a=1...N_{\m{f}}^{(4)}} \,\rightarrow\, \left(\bar{\chi}^i\chi^i\right)^2_{i=1...N_{\m{f}}^{(2)}}, \qquad \left(\bar{\psi}^a\gamma_\mu\psi^a\right)^2_{a=1...N_{\m{f}}^{(4)}} \,\rightarrow\, \left(\bar{\chi}^i\sigma_\mu\chi^i\right)^2_{i=1...N_{\m{f}}^{(2)}}
\label{eq:two-component}
\end{equation}
In the special case of $N_{\m{f}}^{(2)}=1$, it is
\begin{equation}
-3(\bar{\chi}\chi)^2 = (\bar{\chi}\sigma_\mu\chi)^2.
\label{eq:mass}
\end{equation}
There is only one independent fermionic interaction term left. The ansatz for the effective action~\eqref{eq:truncation} then reduces to
\begin{equation}
\label{Gammak2}
\Gamma_k[\bar{\psi},\psi,a] = \int\dif^3{x}\left\{\bar{\chi}\left(iZ_\psi\slashed{\partial} + \bar{e}\slashed{a}\right)\chi + \frac{Z_a}{4}F_{\mu\nu}F_{\mu\nu} + \frac{Z_a}{2\xi}(\partial_\mu a_\mu)^2 + \bar{\lambda}(\bar{\chi}\chi)^2\right\}\;
\end{equation}
with $\bar{\lambda} = \tilde{\bar{g}}_k - 3\bar{g}_k$. Making use of the flow equations~\eqref{4DBetaFuncs}, the $\beta$-function for $\bar{\lambda}$ can be obtained from $\partial_t\bar{\lambda} = \partial_t\tilde{\bar{g}}_k - 3\partial_t\bar{g}_k$. Consequently, the flow equation for the dimensionless renormalized coupling $\lambda = \bar{\lambda}k Z_\psi^{-2}$ is given by
\begin{equation}
\partial_t \lambda = \lambda(1+2\eta_\psi) + 2l^{1,1}_{a,\psi} \lambda e^2 + l^{2,1}_{a,\psi}e^4.
\label{eq:lambda}
\end{equation}
The key feature of this equation is the absence of a $\lambda^2$ term. This is due to 
cancellations in the $\beta$-function special to the $N_{\rm f}^{(2)} = 1$ case as we explain 
in Appendix \ref{FermiSelfApp}. This structural specialty is already 
visible upon setting $N_{\rm f}^{(4)} = 1/2$ in Eq.~(\ref{gBeta}); then the $\tilde{g}^2$ 
in that equation disappears and the fixed-point structure qualitatively changes. 
Further implementing the symmetry Eq.~(\ref{eq:mass}), then yields Eq.~(\ref{eq:lambda}) 
for a single four-fermi coupling. Of course, this $\beta$-function Eq.~(\ref{eq:lambda}) 
may also be derived directly from applying the 
Polchinski-Wetterich equation \cite{polchinski84,WetterichEq} to the ansatz~\eqref{Gammak2}. 
This is performed explicitly in Appendix~\ref{app:betalambda}.

\begin{figure}
\begin{tabular}{ccc}
(a)
&& 
(b)\\
\begin{minipage}{75mm}
\includegraphics[width=75mm]{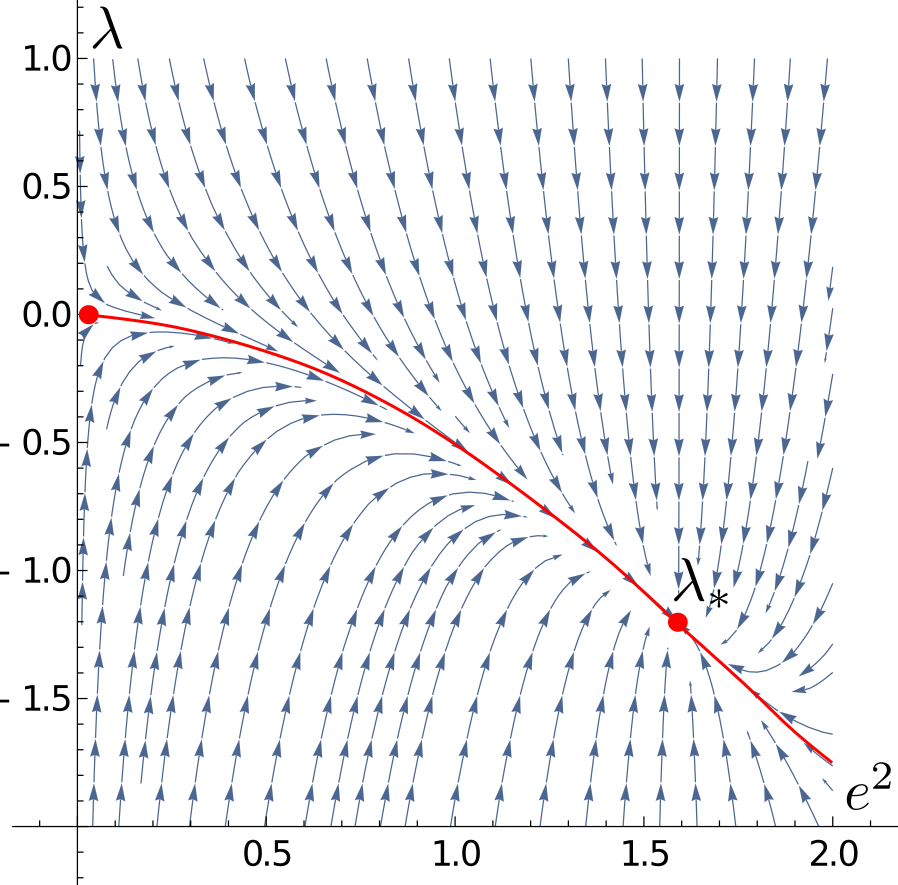}
\end{minipage}
\hspace{5mm}
&&
\begin{minipage}{80mm}
\includegraphics[width=80mm]{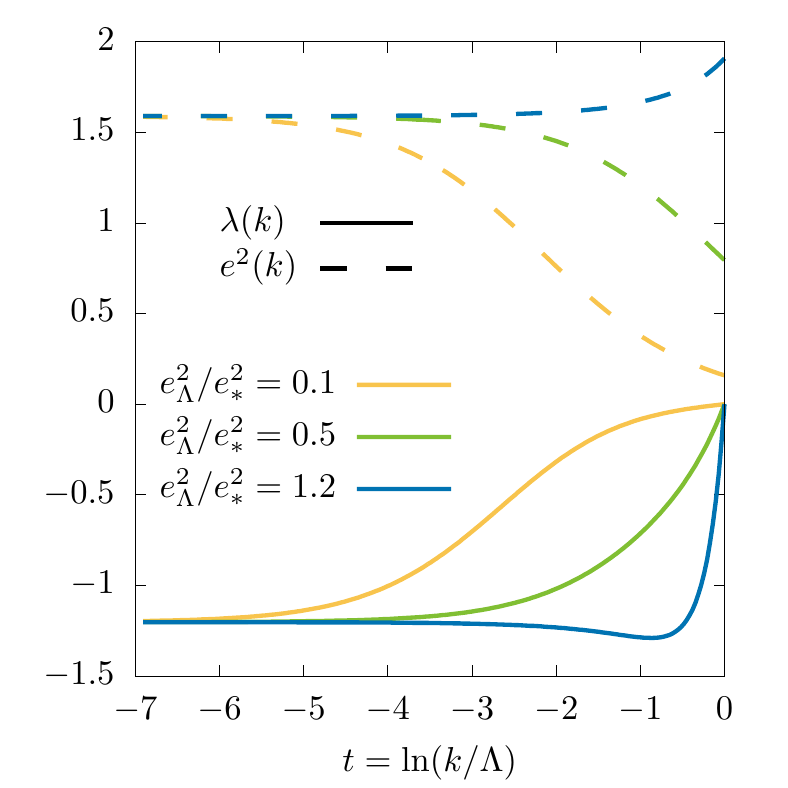}
\end{minipage}
\end{tabular}
\caption{(a): Universal location of the interacting conformal fixed point in the plane of 
interaction couplings for $N_{\m{f}}^{(2)} = 1$. Arrows point towards the infrared. 
(b): Explicit numerical solutions for a range of initial values $e^2_\Lambda$
showing convergence of both, the gauge coupling $e^2$ and the four-fermi coupling $\lambda$, 
toward fixed-point plateaus in the infrared as $k\rightarrow 0$.}
\label{ConfFlow}
\end{figure}

\subsection{Interacting conformal fixed point}
\label{subsec:runaway}
We now adapt the symmetry-breaking analysis explained above 
to a reduced number of equations, 
feeding the gauge sector and anomalous dimensions into Eq.~(\ref{eq:lambda}).
Given the IR fixed point of the gauge sector ($e^2 = e_*^2=1.59$, $\eta_a^\ast = 1$, $\eta^\ast_\psi=-0.64$)
as an input, and because $\partial_t\lambda$ is linear in $\lambda$, there can be only one fixed point solution to Eq.~(\ref{eq:lambda}):
\begin{equation}
\lambda_* = -\frac{l^{2,1}_{A,\psi}e_*^4}{1 + 2\eta_\psi + 2l^{1,1}_{A,\psi}e_*^2} = -1.20, \qquad \pdif{(\partial_t\lambda)}{\lambda}\Bigg|_{(\lambda_*)} = 1+2\eta_\psi+l^{1,1}_{A,\psi}e_*^2 = 3.87
\end{equation}
The numerical values are provided for the linear regulator at $\zeta=0$ and in Landau gauge as before.
Given the positive slope of $\partial_t\lambda$, the fixed point $\lambda_*$ is found to be infrared attractive. 
Therefore, no runaway flow occurs for arbitrary initial values $e^2_{k=\Lambda}$ and 
the fixed-point structure and explicit flows (see Fig.~\ref{ConfFlow}) 
preserve scaling/conformal invariance as $k\rightarrow 0$.
\subsection{Discussion}
In principle, by virtue of Eqs.~(\ref{eq:two-component},\ref{eq:mass}) 
a runaway flow in the four-fermi sector with gauge-invariant regularization 
can lead to quadratic mass terms $\sim m \bar{\chi} \chi$ and 
spontaneous background currents $\sim \langle j^\mu \rangle \bar{\chi} \sigma^\mu \chi$ 
thereby breaking time-reversal and space-reflection symmetry. In fact, this parity anomaly 
appears generically if the entire large group of gauge transformations is allowed and 
Chern-Simons terms are induced \cite{niemi83, redlichprl84,redlichprd84,mulligan13}.
But Eq.~(\ref{eq:model}) with $N^{(2)}_{\rm f} = 1$ is an effective \emph{dual} theory for the 
2D surface of certain topological insulators. 
Then $\chi$ is actually a composite fermion field, which is 
electrically neutral and the standard large gauge transformations 
need to be modified such that no Chern-Simons term from a parity anomaly is allowed 
\cite{metlitski15}.

We note that a  possible ``strong form'' of a Dirac fermion duality 
would relate Eq.~(\ref{eq:model}) with $N^{(2)}_{\rm f} = 1$ to the partition 
function of a single, non-interacting Dirac cone \cite{mross15}.
In this scenario, it can be asserted that if $m_e = 0$ for the 
free Dirac cone, than $m = 0$ for the composite fermions of dual
QED$_3$. Our computation is in line with this reasoning.

\section{Implications of a possible ``strong'' Dirac fermion duality}
\label{sec:strong}
In this section, and based on the considerations above, 
we want to take for granted 
this strong duality on the level of the path integral 
between Eq.~(\ref{eq:lagrangian}) 
and a free Dirac cone of electrically charged fermions $\bar{\psi}_e$, $\psi_e$ 
\cite{mross15}. For a given physical electromagnetic field $A_{\mu}$ coupling the 
electric charge, we have that 
\begin{align}
\mathcal{L}_{free}= \bar{\psi}_e
\left[i\gamma^\mu (\partial_\mu - i e_{\rm phys}  A_\mu)\right]\psi_e\;,
\label{eq:electron}
\end{align}
is ``dual'' to the theory
\begin{align}
\mathcal{L}_{cf}= \bar{\psi}
\left[i\gamma^\mu (\partial_\mu - i g_{cf} a_\mu)\right]\psi+ 
\frac{g_{cf}}{4\pi} \epsilon^{\mu\nu\lambda}A_{\mu}\partial_{\nu}a_{\lambda}\;.
\label{eq:compfermion}
\end{align} 
The duality then implies that both Eq.~(\ref{eq:lagrangian}) and 
Eq.~(\ref{eq:electron}) describe the same underlying physical system 
and the response to actual physical electromagnetic 
fields must be the same. In Eq.~(\ref{eq:electron}), 
the physical electromagnetic field couples, as usual, to the 
electronic current, whereas in Eq.~(\ref{eq:compfermion}) $A_\mu$ 
couples to the gauge flux of the emergent photon.

We now survey two aspects of this duality: constraints for the topological 
current correlator in Subsec.~\ref{subsec:topo} and 
operator dimensions in Subsec.~\ref{subsec:mass}.

\subsection{Constraining the topological current correlator}
\label{subsec:topo}

The current-current correlator on the free Dirac cone side 
of the duality Eq.~\eqref{eq:electron} is
\begin{align}
\langle j_\mu(-p) j_\nu (p) \rangle = - N^{(2)}_{\rm f} C_j^{free} |p| \left(\delta_{\mu \nu} - \frac{p_\mu p_\nu}{p^2}\right)
\label{eq:C_J}
\end{align}
where $j_\mu = \bar{\psi}_e \gamma_\mu \psi_e$ is the physical current dual to the 
physical electromagnetic gauge field $A_\mu$, 
$N^{(2)}_{\rm f} = 1$ for one single two-component fermion, and 
\begin{align}
C_j^{free}=\frac{e_{\rm phys}^2}{16}\;.
\label{eq:exact}
\end{align}
This equation is exact; there are no interaction corrections.

On the composite fermion (cf) side, the physical electromagnetic gauge field couples (via 
a $A_\mu J^{\rm top}_\mu$ term in the Lagrangian)
to the topological current via 
\begin{align}
J^{\rm top}_\mu = \frac{g_{cf}}{4\pi} \epsilon_{\mu}^{\nu\kappa} \partial_{\nu} a_\kappa\;.
\label{eq:topo}
\end{align}
Since both sides of the duality should describe the same physical reality, we should have 
\begin{align}
\langle j_\mu(-p) j_\nu (p) \rangle_{free} = \langle J^{\rm top}_\mu(-p) J^{\rm top}_\nu(p) \rangle_{cf}\, 
\label{eq:current_matching}
\end{align}
%
%
Now, for zero doping and in a conformal phase with conserved topological current 
$\partial_\mu J^{\rm top}_\mu = 0$, the correlator must also have the form
of Eq.~(\ref{eq:C_J}) but with an independent universal constant
\begin{align}
\langle J^{\rm top}_\mu(-p) J^{\rm top}_\nu (p) \rangle = - N^{(2)}_{cf} C_J^{\rm top } 
|p| \left(\delta_{\mu \nu} - \frac{p_\mu p_\nu}{p^2}\right)
\end{align}
with $N^{(2)}_{cf} = 1$. However, the number $C_J^{\rm top}$ is not known, since the composite fermion 
theory is interacting. We can now use Eq.~(\ref{eq:current_matching}) 
to determine $C_J^{\rm top}$ {\it exactly} thus constraining perturbative computations for $C_J^{\rm top}$.
The best estimate for $C_J^{\rm top}$ is in Eq. (4.3.) of Ref.~\cite{GTK16}, abbreviated as GTK:
\begin{align}
\langle J^{\rm top, GTK}_\mu(-p) J^{\rm top, GTK}_\nu (p) \rangle 
&=
- \frac{ 8 |p|}{\pi^2 N_{GTK}}
\left(1 + \frac{1}{N_{GTK}}
\left(
8 - \frac{ 736}{9 \pi^2} \right)
+ 
O(1/N^2_{GTK})
\right)
\left(\delta_{\mu \nu} - \frac{p_\mu p_\nu}{p^2}\right)
\\
&\equiv
-|p|
\left[
\frac{ 8}{\pi^2 N_{GTK}}
\left(1 + \frac{1}{N_{GTK}}
\left(
8 - \frac{ 736}{9 \pi^2} \right)
\right)
+
\Delta X^{\rm top}(N_{GTK})
\right]
\left(\delta_{\mu \nu} - \frac{p_\mu p_\nu}{p^2}\right)\;,
\nonumber
\end{align}
where in the second line we have the subsumed 
the unknown interaction corrections to all 
orders in $1/N_{GTK}$ into the variable $\Delta X^{\rm top}(N_{GTK})$.
After matching conventions
 \begin{align}
  N_{GTK}&=2 N_{\rm f}^{(2)} 
  \nonumber\\
  J_\mu^{\rm top, GTK} &= 2 J_\mu^{\rm top}
\end{align}
and invoking Eqs.~(\ref{eq:exact},\ref{eq:current_matching}), 
we obtain for $N^{(2)}_{\rm f} = 1$ (in units where coupling constants are unity)
%
 %
%
%
%
\begin{align}
\Delta X^{\rm top}(N_{GTK}=2)=-0.0973652\;.
\end{align}
Knowledge of the exact value for this and other 
universal constants may help to constrain 
perturbative computations of thermodynamics, entanglement, and 
response functions of interacting, dual QED$_3$ and possibly 
extensions thereof.

\subsection{Scaling dimension of composite mass operator at one-loop}
\label{subsec:mass}
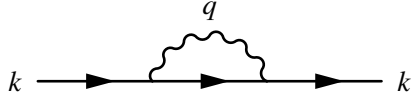
\begin{figure}
\begin{tabular}{ccc}
\begin{fmffile}{diagram2}
\begin{fmfgraph*}(130,82)
\fmfleft{v1}
\fmfright{v2}
\fmf{fermion}{v1,o1,o2,v2}
\fmf{photon,left=0.8,tension=0,label=$q$}{o1,o2}
\fmflabel{$k$}{v1}
\fmflabel{$k$}{v2}
\end{fmfgraph*}
\end{fmffile}
\end{tabular}
\vspace{-10mm}
\caption{One-loop fermion self energy for 
$Z_\psi$ and the corresponding anomalous dimension $\eta_\psi$.}
\label{fig:self}
\end{figure}

Here we compute the singular corrections to the scaling dimension of the 
composite mass operator $m = \bar{\psi} \psi$ for the fermion fields
at the fixed-point of Subsec.~\ref{subsec:runaway}:
\begin{align}
\int d^3 x\, m\, Z_m Z_\psi \, \bar{\psi} \psi
\end{align}
Defining the anomalous exponents ($\Lambda$ is the running cutoff scale),
\begin{align}
\eta_\psi = - \frac{\partial \log Z_\psi}{\partial \log \Lambda}\;,\;\;\;\;\;\;\;\;\;
\eta_m    = - \frac{\partial \log Z_m}{\partial \log \Lambda}\;,
\end{align}
the total correction to the scaling dimension of the mass operator is then
\begin{align}
\eta_{\rm mass} = \eta_m + \eta_\psi\;.
\label{eq:total_mass}
\end{align}
We first compute the fermionic field renormalization 
$Z_\psi$ and $\eta_\psi$ from the one-loop self-energy shown 
in Fig.~\ref{fig:self}. The photon anomalous dimension is $\eta_a = 1$ and we can use the 
standard overdamped one-loop form for $N^{(2)}_{\rm f} = 1$:
\begin{align}
D_{\mu\nu}(q)=\frac{16}{g_{\rm cf}^2|q|}\left(\eta_{\mu\nu}-(1-\xi)\frac{q_{\mu}q_{\nu}}{q^2}\right)\;,
\end{align}
where $g_{\rm cf}$ is the photon-fermion coupling as denoted 
in Eq.~(\ref{eq:compfermion}). 
We may use Feynman gauge $\xi = 1$ in what follows.
%
%
With this, the singular fermion self-energy correction is:
\begin{align}
\delta \Sigma_{\psi}(q)&= g_{\rm cf}^2 \int \frac{d^3 p}{(2\pi)^3}
(-\gamma^\mu)
G_\psi (k+q) 
(-\gamma^\nu) D_{\mu\nu}(q)
\nonumber\\
&=
-\frac{8}{3\pi^2}\slashed{k} \log\left(\frac{\Lambda}{|k|}\right)\;.
\end{align}
Invoking the usual RG improvement, the 
anomalous dimension for the electrons obtains as
\begin{align}
\eta_\psi = \frac{8}{3 \pi^2}\;,
\end{align}
in agreement with a previous calculation upon setting 
their $N$ to $1/2$ and gauge fixing $\xi = 1$ \cite{franz02}.

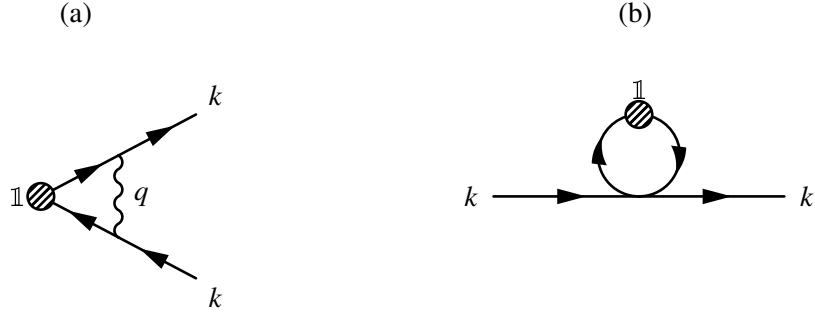
\begin{figure}
\begin{tabular}{ccc}
(a)& &(b)\\[10mm]
\begin{fmffile}{diagram3}
\begin{fmfgraph*}(110,62)
\fmfleft{i1,i2}
\fmfright{v2,v3}
\fmf{fermion}{v2,o1,v1,o2,v3}
\fmf{phantom}{i1,v1,i2}
\fmf{photon,tension=0,label=$q$}{o1,o2}
\fmflabel{$k$}{v2}
\fmflabel{$k$}{v3}
\fmflabel{$\mathbbm{1}$}{v1}
\fmfv{decor.shape=circle,decor.filled=shaded,decor.size=10}{v1}
\end{fmfgraph*}
\end{fmffile}
& \hspace{3cm} &
\begin{fmffile}{diagram4}
\begin{fmfgraph*}(110,62)
\fmftop{v1}
\fmfleft{v2}
\fmfright{v3}
\fmf{fermion}{v2,o1,v3}
\fmf{fermion,left=1,tension=0}{o1,v1,o1}
\fmflabel{$k$}{v2}
\fmflabel{$k$}{v3}
\fmflabel{$\mathbbm{1}$}{v1}
\fmfv{decor.shape=circle,decor.filled=shaded,decor.size=10}{v1}
\end{fmfgraph*}
\end{fmffile}
\end{tabular}\\[5mm]
\caption{One-loop graphs for insertions of the mass operator $m = \bar{\psi} \psi$. (a) is computed 
in the text. (b) arises 
from the four-fermi coupling $\lambda$ and does not produce singular corrections, when approximating 
$\lambda$ as momentum-independent. Upon keeping the momentum-dependence
of the four-fermi coupling, for example via $\sigma$-meson exchange \cite{gross74}, 
(b) can also generate singular corrections to operator dimensions.
}
\label{fig:mass}
\end{figure}

The one-loop graphs for the correction to the mass operator $\delta m(k)$ 
are shown in Fig.~\ref{fig:mass}. 
Graph (a) is non-vanishing and yields
the divergent correction
\begin{align}
\delta m (k)=\mathbbm{1}\frac{24}{\pi^2}\log\left(\frac{\Lambda}{|k|}\right)
\;,\;\;\;\;\;
\eta_m &= - \frac{24}{\pi^2}\;.
\end{align}
Adding the exponents as per Eq.~(\ref{eq:total_mass}) we get the final result
\begin{align}
\eta_{\rm mass} = - \frac{64}{3\pi^2}\;,
\label{eq:mass_final}
\end{align}
which agrees with the leading term in a computation of the same quantity 
by Gracey upon setting his $N_f = 1/2$ \cite{gracey93}.

A strong form Dirac fermion duality would require $\eta_{\rm mass} = 0$ \cite{mross15}.
Implementing the momentum-dependent four-fermi coupling
via $\sigma$-meson exchange \cite{gross74} will produce corrections 
to Eq.~(\ref{eq:mass_final}), as may other additional terms to the truncation 
Eq.~(\ref{Gammak2}). It is an interesting project to establish explicitly
the exponent identities conjectured by the duality.

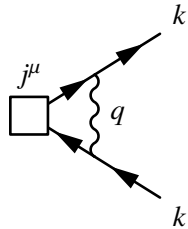
\begin{figure}[b]
\vspace{10mm}
\begin{tabular}{ccc}
\begin{fmffile}{diagram5}
\begin{fmfgraph*}(110,62)
\fmfleft{i1}
\fmfright{v2,v3}
\fmf{fermion}{v2,o1,v1,o2,v3}
\fmf{phantom}{i1,v1}
\fmf{photon,tension=0,label=$q$}{o1,o2}
\fmflabel{$k$}{v2}
\fmflabel{$k$}{v3}
\fmfv{label=$j^{\mu}$,decor.shape=square,decor.filled=empty,label.angle=90,label.dist=10}{v1}
\end{fmfgraph*}
\end{fmffile}
\end{tabular}
\vspace{3mm}
\caption{One-loop correction to insertions of the current operator $j_\mu = \bar{\psi}_{\rm cf} \gamma_\mu \psi_{\rm cf}$.}
\label{fig:current}
\end{figure}

We checked that for the insertion of the (conserved) current operator, 
$ \int d^3 x\, j_\mu \, Z_{j_\mu} Z_\psi \, \bar{\psi}_{\rm cf} \gamma^\mu \psi_{\rm cf}$,
with $j_\mu = \bar{\psi}_{\rm cf} \gamma_\mu \psi_{\rm cf}$, 
these cancellations appear explicitly. 
\begin{align}
\eta_{\rm current} = \eta_{j_\mu} + \eta_\psi =-\frac{8}{3\pi^2}+\frac{8}{3\pi^2} =0\;.
\label{eq:current_cancel}
\end{align}
Evaluating Fig.~\ref{fig:current} results in
\begin{align}
\delta j^{\mu}(k)&=\frac{8}{3\pi^2}
\left(\gamma^{\mu}\log\left(\frac{\Lambda}{|k|}\right)-\frac{k^{\mu}}{k^2}\slashed{k}\right),
\end{align}
and produces a singular correction with opposite sign $\eta_{j_\mu} = - \eta_\psi$ ensuring 
the cancellation Eq.~(\ref{eq:current_cancel}).

\section{Conclusions}
\label{sec:conclu}
This paper explored aspects of a particle-vortex duality 
for Dirac fermions in two space dimensions. We raised the question 
whether a single two-component Dirac fermion coupled to 
$U(1)$ gauge field can have a conformally invariant ground state and 
found indications that this may the case, at least for the 
initially only weakly coupled model. The CFT we found is not free 
but has non-trivial anomalous dimensions and finite four-fermi 
couplings. It would desirable to have a (non-perturbative) 
proof of the absence of spontaneous symmetry-breaking in the 
ground state of Eq.~(\ref{eq:lagrangian}) by generalizing the 
Vafa-Witten theorem to smaller flavor numbers down to $N_f^{(2)} = 1$.

A strong-form duality between dual QED$_3$ and a free Dirac cone 
would constrain operator dimensions and, as we tried to show, also 
universal constants of electromagnetic response functions.

In the future, it will be interesting to establish exponent identities 
and possibly emergent conservation laws of dual QED$_3$ 
more completely and to higher loop order.
To strengthen the link to a specific condensed matter situation, 
a further topic of interest are the ``non-universalities" of dual QED$_3$ 
such as the kinematics, velocities, additional interactions, 
and energy scales below which the continuum field theories for the 
effective degrees of freedoms emerge.

\acknowledgments
We thank A. Altland, J. Braun, S. Diehl and A. Rosch for fruitful discussions. We also thank 
H. Gies, I. Klebanov and M. Metlitski for insightful correspondences. This work 
was supported by the Leibniz Prize of A. Rosch and by the German Research Foundation (DFG) 
through the Institutional Strategy of the University of Cologne within the 
German Excellence Initiative (ZUK 81).

\bibliographystyle{ieeetr}
\bibliography{20160608dualQED3JHEPv3}

\appendix
\section{Derivation of $\beta$-function $\partial_t \lambda$ for four-fermi coupling $\lambda$}
\label{app:betalambda}
The beta function for the single fermionic coupling $\lambda$ can be derived without making reference to the flow equations~\eqref{4DBetaFuncs}. Here we employ the Polchinski-Wetterich equation~\cite{WetterichEq}
\begin{equation}
\label{WetterichEq}
\partial_t\Gamma_k = \frac{1}{2}\m{STr}\left[\frac{\partial_t R_k}{\Gamma_k^{(2)} + R_k}\right] = \frac{1}{2}\m{STr}\left[\tilde{\partial}_t\mathcal{P}_k\right] + \frac{1}{2}\m{STr}\left[\tilde{\partial}_t\sum_{n=1}^\infty\frac{(-1)^{n-1}}{n}(\mathcal{P}_k^{-1}\mathcal{F}_k)^n\right]
\end{equation}
to extract $\partial_t \lambda$ from the ansatz~\eqref{Gammak2}. The expansion of this equation on terms of propagator and fluctuation matrices $\Gamma_k^{(2)} + R_k = \mathcal{P}_k^{-1} + \mathcal{F}_k$ facilitates a projection onto the respective operator structures. The propagator matrix is given by
\begin{equation}
\label{PropMat}
\mathcal{P}_k^{-1} = \delta^{(3)}(p-q)\begin{pmatrix}P_a^{\mu\nu} & 0 & 0 \\ 0 & 0 & -P_\chi \\ 0 & -P_\chi^T & 0 \end{pmatrix}, \quad P_\chi = \slashed{q}\cdot[Z_\psi(1+r_\psi)]^{-1}, \quad P_a = [\delta_{\mu\nu}q^2-q_\mu q_\nu]\cdot[Z_aq^2(1+r_a)]^{-1}.
\end{equation}
Here, only terms $\sim(\bar{\chi}\chi)$ are of interest. Therefore, any explicit dependence of the fluctuation matrix on the gauge field $a_\mu$ can safely be dropped from the outset and we get
\begin{equation}
\mathcal{F}_k = \begin{pmatrix} 0 & \bar{e}\bar{\chi}_{q-p}\sigma_\mu & -\bar{e}\chi^T_{p-q}\sigma_\mu^T \\
-\bar{e}\sigma_\nu^T\bar{\chi}^T_{q-p} & -2\bar{\lambda}\int_{p_1}\bar{\chi}^T_{p_1}\bar{\chi}_{p-q+p_1} & 2\bar{\lambda}\int_{p_1}\left[\bar{\chi}^T_{p_1}\chi^T_{p-q+p_1} - (\bar{\chi}_{p_1}\chi_{p-q+p_1})\mathbbm{1} \right]\\
\bar{e}\sigma_\nu\chi_{p-q} & 2\bar{\lambda}\int_{p_1}\left[(\bar{\chi}_{p_1}\chi_{p-q+p_1})\mathbbm{1} + \chi_{p_1}\bar{\chi}_{p-q+p_1}\right] & -2\bar{\lambda}\int_{p_1}\chi_{p_1}\chi^T_{p-q-p_1} \end{pmatrix}
\end{equation}
Projecting onto spatially constant fermion fields $\chi_p := \chi\delta^{(3)}(p)$, the basic building block of the expansion~\eqref{WetterichEq} can be expressed as
\begin{equation}
\label{PFBlock}
\left[\mathcal{P}^{-1}_k\mathcal{F}_k\right]_{\mu\nu} = \delta^{(3)}(p-q)\begin{pmatrix} 0 & P_a^{\mu\nu}\bar{e}\bar{\chi}\sigma_\nu & -P_a^{\mu\nu}\bar{e}\chi^T\sigma_\nu^T \\
-P_\chi\bar{e}\sigma_\mu\chi & -P_\chi 2\bar{\lambda}\left[(\bar{\chi}\chi)\mathbbm{1} + \chi\bar{\chi}\right] & P_\chi 2\bar{\lambda}\chi\chi^T\\
P_\chi^T\bar{e}\sigma_\nu^T\bar{\chi}^T & P_\chi^T 2\bar{\lambda}\bar{\chi}^T\bar{\chi} & -P_\chi^T 2\bar{\lambda}\left[\bar{\chi}^T\chi^T - (\bar{\chi}\chi)\mathbbm{1} \right] \end{pmatrix}.
\end{equation}
There are three contributions to the flow of the bare coupling $\bar{\lambda}$ which can be depicted diagrammatically as in fig.~\ref{FermiFlowDiags} below.

Here, double lines symbolize the (full) renormalized propagators in~\eqref{PropMat}.

\paragraph{Fermionic self interaction}
This contribution does not involve the gauge vertex. Consequently, only the lower right submatrix of eq.~\eqref{PFBlock} is needed. Projecting onto $(\bar{\chi}\chi)^2$ gives
\begin{equation}
\partial_t \bar{\lambda}|_{\bar{\lambda}^2} (\bar{\chi}\chi)^2 = \frac{1}{\Omega}\frac{1}{2}\frac{-1}{2}\m{STr}\left[\left(\mathcal{P}^{-1}_k\mathcal{F}_k\right)^2_{\bar{\lambda}^2}\right] = 0
\end{equation}
where $\Omega$ is the three-dimensional spacetime volume. Thus, tracelessness of the quadratic term in the expansion enforces linearity of the $\beta$-function. This finding is crucial for QED$_3$ being conformal for $N_{\m{f}}^{(2)} = 1$ and it only occurs for this flavor number. Some details of why this is the case are given in App.~\ref{FermiSelfApp} below.

\begin{figure}
\centering
\includegraphics[width=15cm]{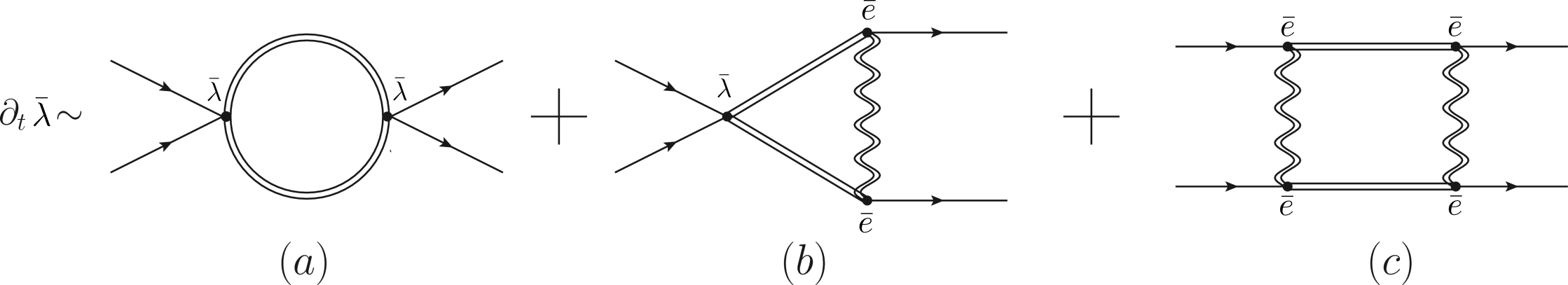}
\caption{Diagrammatic representation of the contributions to $\partial_t \bar{\lambda}$. It 
is a special feature here that diagram (a) vanishes after the trace.}
\label{FermiFlowDiags}
\end{figure}

\paragraph{Triangle diagram}
Both gauge and fermionic sectors are involved in the computation of this diagram. It is given by
\begin{equation}
\partial_t \bar{\lambda}|_{\bar{\lambda}\bar{e}^2} (\bar{\chi}\chi)^2 = \frac{1}{\Omega}\frac{1}{2}\frac{1}{3}\m{STr}\left[\left(\mathcal{P}^{-1}_k\mathcal{F}_k\right)^3_{\bar{\lambda}\bar{e}^2}\right] = \frac{l^{1,1}_{a,\psi}}{\pi^2}\frac{2}{kZ_\psi^2Z_a}(\bar{\chi}\chi)^2
\end{equation}
Here, the usual compact notation in terms of a threshold function has been introduced,
\begin{equation}
l_{a,\psi}^{n_a,n_\psi} = k^{2n_A-2n_\psi-3}\int\dif{q}q^4\left[n_a\frac{\partial_t r_a - \eta_a r_a}{P_{r_a}} + 2n_\psi \frac{1+r_\psi}{P_{r_\psi}}(\partial_t r_\psi - \eta_\psi r_\psi)\right]P_{r_a}^{-n_a}P_{r_\psi}^{-n_\psi},
\end{equation}
with
\begin{equation}
P_{r_a} = q^2\left[1+r_a\right], \quad P_{r_\psi} = q^2[1+r_\psi]^2.
\end{equation}

\paragraph{Box diagram}
The last diagram is particularly important as it is responsible to generate the fermionic interaction when starting from the QED$_3$ action in the UV, where $\bar{\lambda}_\Lambda = 0$. When computing its value, the lower right submatrix of Eq.\eqref{PFBlock} may be ignored as only the gauge vertex contributes.
\begin{equation}
\partial_t \bar{\lambda}|_{\bar{e}^4} (\bar{\chi}\chi)^2 = \frac{1}{\Omega}\frac{1}{2}\frac{-1}{4}\m{STr}\left[\left(\mathcal{P}^{-1}_k\mathcal{F}_k\right)^4_{\bar{e}^4}\right] = \frac{l^{2,1}_{a,\psi}}{\pi^2}\frac{1}{k^3Z_\psi^2 Z_a^2}\bar{e}^4(\bar{\chi}\chi)^2
\end{equation}

\paragraph{Renormalization and rescaling}
In a last step, a renormalization of the couplings as in eq.~\eqref{RenPres} and the fields as $\chi \,\rightarrow\, \chi/\sqrt{Z_\psi}$ as well as a rescaling of $e^2$ and $\lambda$ with $\pi^2$ provides the final result for the $\beta$-function:
\begin{equation}
\label{FinalBeta}
\partial_t \lambda = \lambda(1+2\eta_\psi) + \frac{k}{Z_\psi^2}\partial_t\bar{\lambda} = \lambda(1+2\eta_\psi) + 2l^{1,1}_{A,\psi} \lambda e^2 + l^{2,1}_{A,\psi}e^4.
\end{equation}
This reproduces Eq.~(\ref{eq:lambda}) of the main text.

\section{Cancellation of $\lambda^2$ term from flavor trace in $\partial_t \lambda$}
\label{FermiSelfApp}

In order to understand the origin of the cancellation enforcing linearity of the $\beta$-function~\eqref{FinalBeta} with respect to $\lambda$, it is necessary to revisit the corresponding contributions for general flavor number $N_{\m{f}}^{(2)}$. Although there are two independent quartic fermion terms for $N_{\m{f}}^{(2)}>1$ (see Eq.~\eqref{eq:two-component}), it is sufficient to consider the generalized interaction 
\begin{equation}
(\bar{\chi}\chi)^2 \,\rightarrow\,\left(\bar{\chi}^i\chi^i\right)^2_{i=1...N_{\m{f}}^{(2)}}.
\end{equation}
The fermion sector of the analog to eq.~\eqref{PFBlock} is then given by
\begin{equation}
\label{FBlock}
\left[\mathcal{P}^{-1}_k\mathcal{F}_k\right]^{ij}_{\bar{\chi}\chi} = \delta^{(3)}(p-q)\frac{2\bar{\lambda}}{N_{\m{f}}^{(2)}}\begin{pmatrix}
-P_\chi \left[(\bar{\chi}\chi)\mathbbm{1}\delta^{ij} + \chi^i\bar{\chi^j}\right] & P_\chi \chi^i\left(\chi^j\right)^T\\
P_\chi^T \left(\bar{\chi}^i\right)^T\bar{\chi}^j & -P_\chi^T \left[\left(\bar{\chi}^i\right)^T\left(\chi^j\right)^T - (\bar{\chi}\chi)\mathbbm{1}\delta^{ij} \right] \end{pmatrix},
\end{equation}
where the flavor dependence of the fermion propagator has already been absorbed. It is then
\begin{equation}
\begin{aligned}
\partial_t \bar{\lambda}|_{\bar{\lambda}^2} (\bar{\chi}^i\chi^i)^2 &= \frac{N_{\m{f}}^{(2)}}{\Omega}\frac{1}{2}\left(\frac{-1}{2}\right)\m{STr}\left[\left(\mathcal{P}^{-1}_k\mathcal{F}_k\right)^2_{\bar{\lambda}^2}\right] = 2\frac{l^1_{\psi}}{N_{\m{f}}^{(2)}\pi^2}\m{Tr}_{\m{f}}\left[\left(\bar{\chi}^a\chi^a\right)\bar{\chi}^i\chi^j - \left(\bar{\chi}^a\chi^a\right)^2\delta^{ij}\right]\\
&= 2\frac{l^1_{\psi}}{N_{\m{f}}^{(2)}\pi^2}\left(1-N_{\m{f}}^{(2)}\right)\bar{\lambda}^2\left(\bar{\chi}^i\chi^i\right)^2
\label{eq:cancel}
\end{aligned}
\end{equation}
where in the first line all traces but the flavor one have been performed and
\begin{equation}
l_\psi^n = 2nk^{2n-3}\int\dif{q}q^4(\partial_t r_\psi - \eta_\psi r_\psi)\frac{1+r_\psi}{P_{r_\psi}^{n+1}}.
\end{equation}
We emphasize the prefactor in this expression vanishes for the case 
of interest $N_{\rm f}^{(2)}=1$ as announced in the main text.
This is due to the absence of flavor off-diagonal terms in Eq.~(\ref{eq:cancel}) 
for this case. For $N_{\rm f}^{(2)}=1$, the flavor indices $a=i=j$ must fall onto each other 
and the bracket $[...]$ in the first line of Eq.~(\ref{eq:cancel}) vanishes.
Related vanishing of $\beta$-functions in the single-flavor case 
also appears in the Gross-Neveu model and in particular also holds at 
higher loop orders \cite{wetzel85,gracey90, luperini91}.
Complete cancellations of individual contractions in a $\beta$-function 
for a four-fermion vertex also appear in Luttinger liquids 
\cite{shankar94}, although these are specific to spatial dimension $D=1$.

\end{document}